# Blackbody Friction: Analytic expressions for velocity and position

Joseph West

*Department of Chemistry and Physics, Indiana State University, Terre Haute, IN 47809*

E-mail: joseph.west@indstate.edu

**Abstract.** The equations of motion are solved analytically for speed and position for the case of blackbody friction on objects traveling at relativistic speeds with respect to observers fixed in the frame of the blackbody radiation. Two model cases are considered, both assume an object which is a perfect absorber of light, and one of which also assumes that the object maintains a constant rest mass by emitting photons in a momentum neutral process. The maximum distances attained for a light sail and a micron-size sand grain are determined under the assumption that the relevant blackbody source is that of cosmic background radiation. At a temperature of 3000K, the temperature at the time the universe became transparent, cosmic background radiation would cause a significant decay of speed fluctuations in dust clouds.

## I. INTRODUCTION

In special relativity, there are few instances where a force law allows for an analytical result for the dynamics of an object. The trivial case of no net force, constant velocity, is one example. Another is the case of constant net force (constant as measured by the accelerating observers).[1,2] However, resistive forces which at the microscopic level are the direct result of **collisions** offer the opportunity to be dealt with using conservation of energy and momentum, and do not require the introduction of force carriers or retarded potentials. In this paper, objects interacting with uniform temperature blackbody radiation which is at rest relative to an inertial reference frame are considered. As a direct consequence of the phenomenon of Stellar aberration, there can be only one unique inertial frame for which the radiation will appear isotropic. Once a solution to velocity and position is known in that frame, the usual Lorentz transformations allow the solution to be formulated in any other inertial reference frame.

The Cosmic Microwave Background (CMB), the "remnant of the Big Bang," will be used as an example photon field. It is largely uniform to observers at rest relative to the local Hubble Flow, referred to throughout as the Hubble Flow frame (HFF). The Earth it moves with the local group at a velocity of approximately 600 km/s relative to the HFF, so that the Doppler shift produces a noticeable CMB dipole component.[3] As measured in the HFF, the Earth absorbs the same number of photons as a stationary object, but it absorbs more photons with velocity components directed opposite to its motion than photons moving in the direction of its motion,[4] resulting in a net momentum transfer between the HFF and the Earth, manifesting effectively as a velocity dependent drag force.[5] In the case of isolated atoms or molecules moving within a "thermal photon bath," this effect has been studied in some detail as the Einstein-Hopf drag.[4,6,7,8] In most of that previous work, the details of the frequency distribution of the photon bath and the quantized energy levels of the atom or molecule are important issues, and the primary objective has been to determine an approximate expression for the drag force. Efforts to apply the same ideas to objects traveling relative to the CMB have focused on attempts to determine the form of the drag force, but the author is not aware of any published expressions for velocity or position of those objects as a function of time. For example Mkrtchian et. al. find a force linear in velocity.[8]

In this paper, two idealized macroscopic scale models are presented. Perfect absorption of all photons is assumed for both models, and perfect photon emission is assumed for one of the models. This removes the effects of quantized atomic energy levels, and all details of the frequency distribution of the photon bath. Further, the model is restricted to one dimensional motion for the objects and the photons. For the perfect emitting model, a full three-dimensional photon bath gives the same qualitative behavior, only changing the time scale by a constant factor. To aid in discussion, the objects will be referred to as "ships," with fictitious "crews" to fill the role of "observers."

It is found that there is a characteristic time associated with the motion as described by each model determined by: the CMB intensity; the cross-sectional area of the object along its direction of motion; and the object's initial rest mass. This characteristic time is estimated for object parameters appropriate for a light sail and a micron-size dust grain. For the current CMB intensity, the effects of the blackbody friction are negligible, but for thermal background radiation intensities equivalent to the time at which the universe became transparent, the characteristic time is of the order of only a few years. The remainder of the paper is organized as follows: section II introduces the models, and the relevant derivations of the speed (IIa) and position (IIb) as functions of time, section III explores the qualitative behavior of the speed and position expressions, and the light sail and sand parameters are considered, and in section IV, the conclusions are presented.

## II. OBJECT AND PHOTON INTERACTIONS

The postulates of special relativity imply that there is no preferred reference frame, that all inertial observers are equivalent. However, for observers (on a "ship") moving relative to the HFF, the CMB has a dipole signature due to a relativistic Doppler shift as the universe as a whole moves relative to the ship. In addition to the Doppler shift of the photon energy, a well-known aspect of stellar aberration is that the **number** of photons incident on the ship is also affected.[9] More photons, all blue shifted, are incident on the "front" of the ship. Fewer photons, all red-shifted, are incident on the "back" of the ship which are all red-shifted.[2,10] The crew will observe a stellar background with the stars concentrated to the front **and** the rear of the ship, with the effect increasing with increasing relative speed. The apparent redistribution of the photon sources in the crew frame relative to a uniform distribution of stars has been considered by others, with the work by Western and by Lagoute and Davoust being especially instructive.[2,10] This "redistribution" of the photon impacts on the ship to the front and back surfaces partially justifies the use of the one-dimensional model used in this paper.

Two physical models are considered for the ships. Each ship is assumed to be cylindrical in shape, with cross-sectional area A, and rest mass m. The first ship is a perfect absorber of photons of all frequencies, referred to as the ASH (Absorbing SHip).

The total energy of this ship, as well as its rest mass, will increase continuously as photons are absorbed. Presumably, as its rest mass increases, new matter is added to the ship in such a way as to increase its length, while retaining the constant cross-sectional area (and the one-dimensional model).

The second ship is a perfect absorber of photons, but also is assumed to emit photons symmetrically as seen by the crew of the ship (via thermal radiation, or maybe a laser cooling system), such that the rest mass of the ship remains constant. This ship will be referred to at the DEM ("DE-Massing" ship). If photons are emitted by a body symmetrically in that body's rest frame, then it can be shown that all observers agree that the emitted photons have **no** effect on the speed of the ship.[11] The analysis of French of the Photon Rocket, which follows from Pierce but with the derivation run "backwards" provide the necessary arguments.[12,13] In fact, much of the derivation that follows owes a great deal to this example. For wavelengths on the order of the size of the ship, thermal radiation is not emitted symmetrically, and can even be emitted with a dependence on temperature to the 6th power, instead of the usual Stefan-Boltzmann 4th power dependence.[14] Those effects are not included in the model presented here.

**II A. Velocity vs Time**

Consider the dynamics of light absorption of a ship moving to the right (+x direction) as viewed in the HFF. During a short time dt, the ship absorbs an energy $E_R$ from photon traveling **to the right** (in the positive x direction) and an energy $E_L$ from photons traveling to the left (negative x direction). There is a change in momentum of the ship equal to $E_R/c - E_L/c$, where c is the speed of light. The rest mass of the ship at time t is m, and the rest mass at time t + dt is m'. During the time dt, the value of $\beta = v/c$ changes to $\beta' = v'/c$, where v is the velocity of the ship, and the associated relativistic factor changes from $\gamma$ to $\gamma'$, where $\gamma = (1-\beta^2)^{-1}$. Conservation of energy (a) and momentum (b) then gives

$$\gamma' m' c^2 = E_R + E_L + \gamma m c^2 \tag{1a}$$

$$\gamma' m' v' = E_R/c - E_L/c + \gamma m v \tag{1b}$$

Observers in the HFF see a fixed number of photons absorbed during the time dt, **regardless of the motion of the ship**. The ship is effectively "running away" from, and absorbs fewer photons moving to the right, but manages to run into the same number of "extra" photons that are moving to the left. The crew and observers in the HFF must agree about the ratio of the number of photons absorbed from each direction. The crew attributes the asymmetry to an observed difference in the density of photon sources between the front and rear of the ship.[2,15] The amount of energy absorbed depends on the intensity of the CMB, modified by the velocity of the ship relative to the CMB, and the area A, presented to those photons. Observers in the HFF find that

$$E_R = (1-\beta) A \sigma T^4 dt \equiv \lambda(1-\beta)dt \qquad E_L = \lambda(1+\beta)dt, \tag{2}$$

where $\sigma$ is the Stefan-Boltzmann constant, ($\sigma = 5.67 \times 10^{-8}$ W/(m²K⁴)), T is the temperature in Kelvin, and $\lambda$ is the rate in Watts at which the ship would absorb energy at either end when at rest with respect to the HFF. Combining the expressions in Eqs. (1) and (2) gives

$$\beta' = \frac{\beta + R - L}{1 + R + L}, \text{ where } \quad R = \frac{E_R}{\gamma mc^2} \quad \text{and} \quad L = \frac{E_L}{\gamma mc^2} \quad (3)$$

The total rate at which the ASH absorbs energy in the HFF is $2\lambda$, regardless of the motion of the ship (see Eqs. (1) and (2)). The **total** energy of the ASH is therefore known exactly as a function of time in the HFF frame

$$\gamma_{ASH} m_{ASH} c^2 = \gamma_o m_o c^2 + 2\lambda t \equiv E_o + 2\lambda t, \quad (4)$$

where $m_o$, $\lambda_o$, and $E_o$ are the rest mass, relativistic factor, and total energy of the ship respectively at time $t = 0$. Combining Eqs. (3) and (4), and dividing by dt gives the rate of change of $\beta$ for the ASH in the HFF frame:

$$\left(\frac{d\beta}{dt}\right)_{ASH} = -\frac{4\lambda\beta}{E_o + 2\lambda t}. \quad (5)$$

For the DEM, the emission of photons by the ship is used to **force** the rest mass to remain constant, $m = m_o$, so that a simple rearrangement of Eq. (3) and dividing by dt gives

$$\left(\frac{d\beta}{dt}\right)_{DEM} = -\frac{4\lambda\beta\sqrt{1-\beta^2}}{m_o c^2}. \quad (6)$$

Consistent with previous investigators,[4,8,15] for short times (Eq. (5)) and small values of $\beta$ (Eq. (6)), the acceleration and "effective drag force" is linear in $\beta$. For longer times, however, there are significant deviations from that linear behavior. For the DEM it is seen from Eq. (6) that there is in fact a maximum acceleration at an intermediate value of $\beta$ ($\beta = 1/\text{sqrt}(3)$ [$\beta = 0.577$]).

For both models, the time scale of the dynamics of the objects is determined by the photon density and the initial rest mass of the ship. The time scale is defined by

$$t_o \equiv m_o c^2 / \lambda. \quad (7)$$

Using time measured in units of $t_o$, Eq. (5) integrates directly to give[16]

$$\ln\left(\frac{\beta}{\beta_o}\right)_{ASH} = -2\ln(\gamma_o + 2t) + 2\ln(\gamma_o) = \ln\left(\frac{\gamma_o^2}{(\gamma_o + 2t)^2}\right) \quad (8)$$

which can be rearranged to give $\beta$ directly as an analytic function of time

$$\beta_{ASH} = \left(\frac{dx}{dt}\right)_{ASH} = \frac{\gamma_o^2 \beta_o}{(\gamma_o + 2t)^2} \quad (9)$$

Likewise, the DEM is amenable to an analytic solution. Integrating Eq. (6) gives[17]

$$-\ln\left(\frac{1+\sqrt{1-\beta^2}}{\beta}\right) + \ln\left(\frac{1+\sqrt{1-\beta_o^2}}{\beta_o}\right) = -4t \quad (10)$$

Straightforward algebra allows one to obtain $\beta$ analytically in terms of the initial conditions and the time elapsed in the HFF frame,

$$\beta_{DEM} = \left(\frac{dx}{dt}\right)_{DEM} = \frac{2}{De^{4t} + (D^{-1})e^{-4t}}. \qquad D \equiv \frac{1+\sqrt{1-\beta_o^2}}{\beta_o} \quad (11)$$

**II B. Position vs Time**

For both ships, the velocity goes to zero as the time gets "large," implying for both cases the possibility of an upper bound on the total distance traveled. To that end, Eq. (9) can be integrated directly,[18] and setting the initial position to zero gives

$$x_{ASH} = \frac{c\gamma_o \beta_o t}{2t + \gamma_o} \quad (12)$$

The position of the DEM can likewise be obtained in closed form from of Eq. (11) (see note regarding incorrect entry in integral tables)[19] via

$$\int \frac{dt}{Pe^{bt} + Qe^{-bt}} = \frac{1}{b\sqrt{PQ}} \tan^{-1}\left(e^{bt}\sqrt{\frac{P}{Q}}\right). \quad (13)$$

Setting the initial position to zero, Eq. (11) can be integrated using Eq. (13) to give

$$x_{DEM} = \frac{ct_o}{2}\left[\tan^{-1}\left(\frac{e^{4t}}{D}\right) - \tan^{-1}\left(\frac{1}{D}\right)\right] = \frac{c\lambda}{2m_o c^2}\left[\tan^{-1}\left(\frac{e^{4t}}{D}\right) - \tan^{-1}\left(\frac{1}{D}\right)\right] \quad (14)$$

Knowing the velocity and position in the HFF inertial frame allows one to use the usual Lorentz transformations to find the velocity and position of the objects in any other inertial reference frame.

**III. RESULTS**

Graphs indicating the qualitative time dependence of $\beta_{ASH}$, $\beta_{DEM}$, $x_{ASH}$ and $x_{DEM}$ are presented in Figs. 1 and 2. The values for Fig. 1 correspond with a "high" initial speed ($\beta_o = 0.95$), while Fig. 2 illustrates values for a "low" initial speed ($\beta_o = 0.01$). For both choices (Fig. 1a and 2a) $\beta_{ASH} > \beta_{DEM}$ for t > 0. The difference $\beta_{ASH} - \beta_{DEM}$ reaches a maximum (Fig 1b and 2b) at t = 0.57, near the time for the maximum acceleration of the DEM as noted above. The maximum point is also evident as a change in the sign of the curvature of the graph of $\beta_{DEM}$ in Fig. 1c. In contrast for low initial speeds the DEM remains below that critical speed, and the curvature of $\beta_{DEM}$ in Fig. 2c does not change sign.

For t "large," the exponent in Eq. (14) becomes very large, so that the first term of inverse tangent returns a value of $\pi/2$. The motion for the DEM for "large" times therefore approaches a limiting value (see Figs. 1d and 2d)

$$x_{DEM}(t \to \infty) = \frac{ct_o}{2}\left[\frac{\pi}{2} - \tan^{-1}\left(\frac{1+\sqrt{1-\beta_o^2}}{\beta_o}\right)\right] \quad (15)$$

For an initial speed of zero, the argument of the second tangent term also approaches infinity, giving a net displacement of zero, as expected. For an initial speed approaching $\beta_o = 1$ (again, see Fig. 1d), the remaining inverse tangent term in Eq. (15) returns a value of $\pi/4$, so that

$$x_{DEM}(\beta_O \to 1, t \to \infty) = \frac{\pi}{8}ct_o. \quad (16)$$

This indicates that there is an upper limit on the range of the DEM, **regardless of the total initial energy (or initial velocity)**.

In contrast, the total displacement of the ASH increases without bound with increasing values of $\beta_o$

$$x_{ASH}(t \to \infty) = \frac{v_o E_o}{2\lambda} = \frac{\beta_o \gamma_o}{2}ct_o. \quad (17)$$

It should be noted that the ASH model does seem to accumulate an "unreasonable" amount of rest mass. For example, with an initial speed $\beta_o = 0.95$, the ASH would increase its rest mass to almost **five times** its starting value ($m = 5m_o$) by the time $t = t_o$.

Consider the effects of a photon field on two bodies with relatively large area to volume ratios: the Ikaros light sail and a a micron sized grain of sand. The square light sail has a total area $A = 196$ m$^2$, is made from polyimide resin, which is as thin as 0.0076 mm (7.6 x 10$^{-6}$ m) in some spots.[20] The density of the resin is approximately 1.0 kg/m$^3$,[21] so that a lower limit on the rest mass of the sail without a payload is $m_o = 1.45$ x 10$^{-3}$ kg ($m_o c^2 = 1.30$ x 10$^{14}$ J). For comparison, the total mass of the sail and its actual payload is 316 kg. It is assumed that the sand grain has a density of 2 x 10$^3$ kg/m$^3$ and is in the shape of a cylinder of radius and length R, with R= [(1.0 x 10$^{-6}$ m)/$\pi$] giving an area $A = 1.0$ x 10$^{-12}$ m$^2$ and rest mass $m_o = 2$ x 10$^{-15}$ kg ($m_o c^2 = 180$ J).

At the current CMB temperature T = 2.7K, the rate of energy absorbed $\lambda$, by a given area element of area A, is

$$\lambda = A\sigma T^4 = \left(3.01 \times 10^{-6} \frac{J}{sm^2}\right)A \quad (18)$$

For the light sail this gives $\lambda = 5.90$ x 10$^{-4}$ J/s, and $t_o = 2.21$ x 10$^{17}$ s (7.01 x 10$^9$ yr), which is roughly half the age of the universe. For the DEM, Eq. (16) places an absolute upper limit on the Light Sail displacement of roughly 10$^{12}$ light years (ignoring effects due to the expansion of the universe). For the grain of sand, $t_o = 5.98$ x 10$^{19}$ s (1.90 x 10$^{12}$ yr), much longer than the age of the universe. Clearly the blackbody friction does not produce a significant effect on the motion of these objects under the current conditions of the CMB.

At earlier times, the universe temperature was much higher. At the time the universe became transparent, the temperature was approximately T = 3000K, which reduces $t_o$ by a **factor** of approximately $10^{12}$. For the grain of sand, this higher temperature gives $t_o$ = 2.52 yr, indicating that blackbody friction could have had a strong damping effect on the motion of micron sized particles relative to the local Hubble flow. For the light sail, the time scale is quite short at $t_o = 7 \times 10^{-3}$ years, or about 2.5 days.

The time dependence of $\gamma$ is shown in Fig. 1e. It is seen that for an initial speed $\beta_o = 0.95$, the values of $\beta$ and $\gamma$ both decrease dramatically for both models on the time scale of $t_o$. As a result, time dilation effects for the crew are important only up to a time of approximately $t = t_o$. With an initial speed $\beta_o = 0.99$, the crew elapsed time t' as measured on the ASH, is determined numerically to be t' = $0.58 t_o$ when $t = t_o$. Beyond that time, there is almost no aging difference between the crew and the HFF observers.

## IV. DISCUSSION

The effects of blackbody friction (Einstein-Hopf drag) on objects traveling at relativistic speeds have been examined. Two model cases were considered, the first an object which is a perfect absorber (the ASH), and the second which is a perfect emitter **and** absorber (the DEM). These models result in **analytic** expressions for the velocity and position of the model objects as functions of time in the HFF frame. Analytic expressions for velocity and position are therefore known in any inertial reference frame through the use of the usual Lorentz transformations. Analytic solutions for velocity and position are uncommon, even for force models with the simplest mathematical form, in special relativity. The results are applicable to objects moving relative to a photon background that appears isotropic in one particular inertial reference frame. The Doppler effect ensures that a photon field can be isotropic in only one such inertial frame, but as noted, the results can then be formulated in inertial frames in motion relative to that photon-isotropic frame.

For the same initial speed and mass, it is found that for any given time, the ASH is always traveling faster than the DEM. As expected, an increase in the initial velocity of either object increases its range, both objects have a **finite** range for any given finite initial energy. For the DEM it is found that the range approaches an asymptotic **upper limit** as the energy is increased. The upper limit is determined by the cross-sectional are of the DEM and the overall intensity of the photon bath.

The effects of the blackbody friction due to the CMB on a light sail, and a micron size grain of sand were considered at the current background temperature of 2.7K. The effects of the CMB on the motion of those objects was found to be insignificant for practical applications. However, at a temperature of 3000K, the temperature at the time the universe became transparent, the effects are rather dramatic. Under those conditions, the speed of a micron sized sand grain relative to the CMB would be reduced from 0.95c to well below 0.05c in just a few years, causing a significant decay of speed fluctuations in dust clouds in the early universe.

## ACKNOWLEDGEMENTS

The author would like to extend his thanks to Dan Moser (class of 2012, department of chemistry and physics ISU) for help in confirming the error in the integral table from Schaum's Outlines. See reference 19.

# FIGURE CAPTIONS

Fig. 1. Initial speed $\beta_o = 0.95$. This value is chosen so that the general features of the relationships are evident. Time is measured in units of $t_o = m_o c^2/\lambda$.

Fig1a. Speed as a function of time for the ship which absorbs all energy falling on it (top line in diamonds) and the ship which de-masses (bottom line, squares).

Fig1b. the difference in speed between the two ships. Notice that the fully absorbing ship is always traveling faster.

Fig. 1c. The speed of the de-massing ship, in order to show the changes in curvature as a function of time at this relatively large initial speed. The changes are due to the comments following Eq. (9) with the effects of the resistive photons being largest when $\beta$ is at extreme values.

Fig1d. The position of the ships as a function of time. Absorbing ship data shown as solid diamonds, and De-massing ship data shown in solid squares. Notice how the de-massing ship reaches the upper limit of its motion in a relatively short amount of time.

Fig. 1e. The relativistic factor $\gamma$, for the ASH as a function of time. The ASH is the higher speed case, but even for a relatively high initial speed of $\beta_o = 0.95$, the time dilation effects for events occurring on the ASH are minimal beyond $t = t_o$.

Fig. 2. Initial speed $\beta_o = 0.01$. This value is chosen so that the general features of the relationships are evident. Time is measured in units of $t_o = m_o c^2/\lambda$.

Fig 2a. Speed as a function of time for the ship which absorbs all energy falling on it (top line in diamonds) and the ship which de-masses (bottom line, squares).

Fig 2b. the difference in speed between the two ships. Notice that the fully absorbing ship (top line, blue diamonds) is always traveling faster than the de-massing ship (bottom line, red squares).

Fig. 2c. The speed of the de-massing ship, notice that the curvature remains positive over the entire range.

Fig 2d. The position of the absorbing ship (top line, blue diamonds) and the de-massing ship (bottom line, red squares) as a function of time. Notice that the de-massing ship reaches the upper limit of its motion in a relatively short amount of time at approximately $t = t_o$.

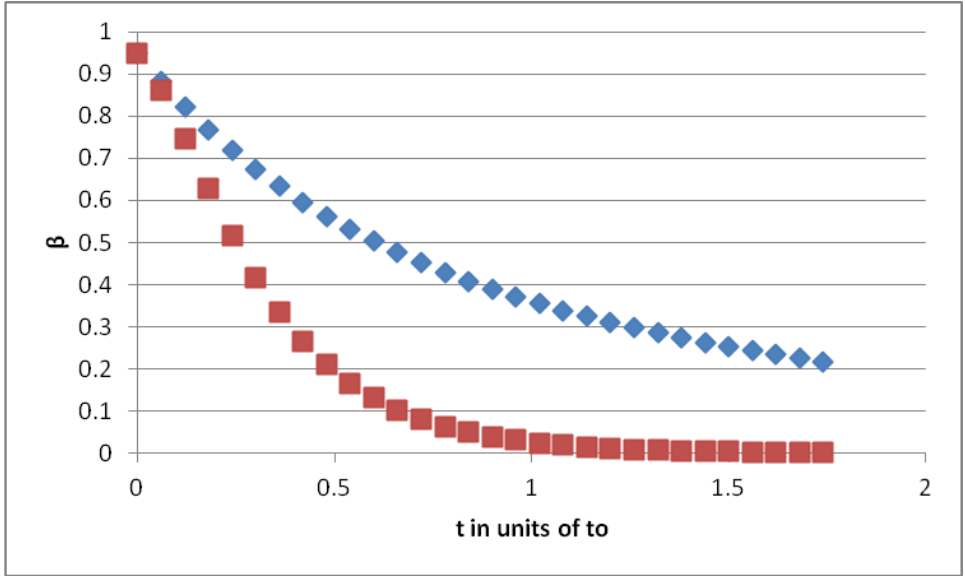

1a

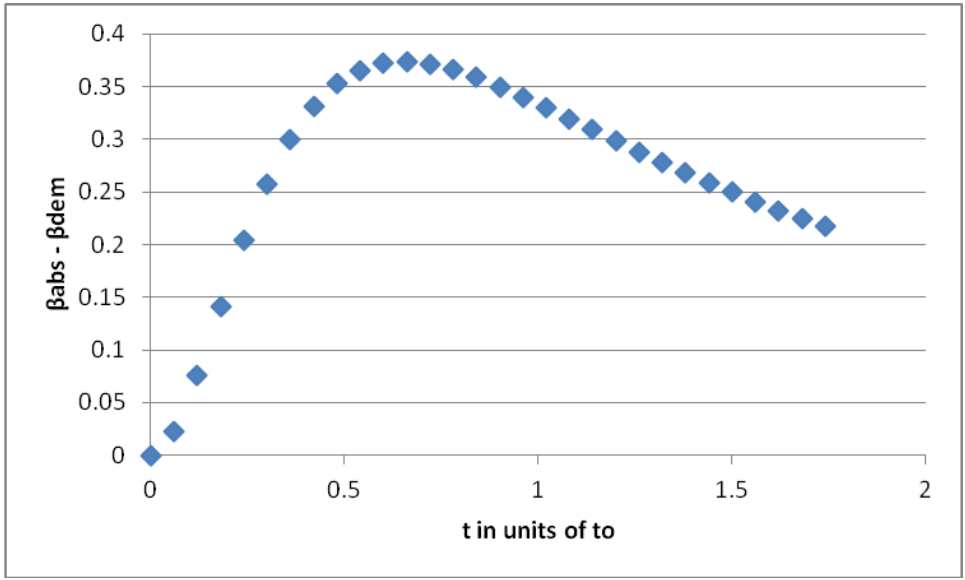

1b

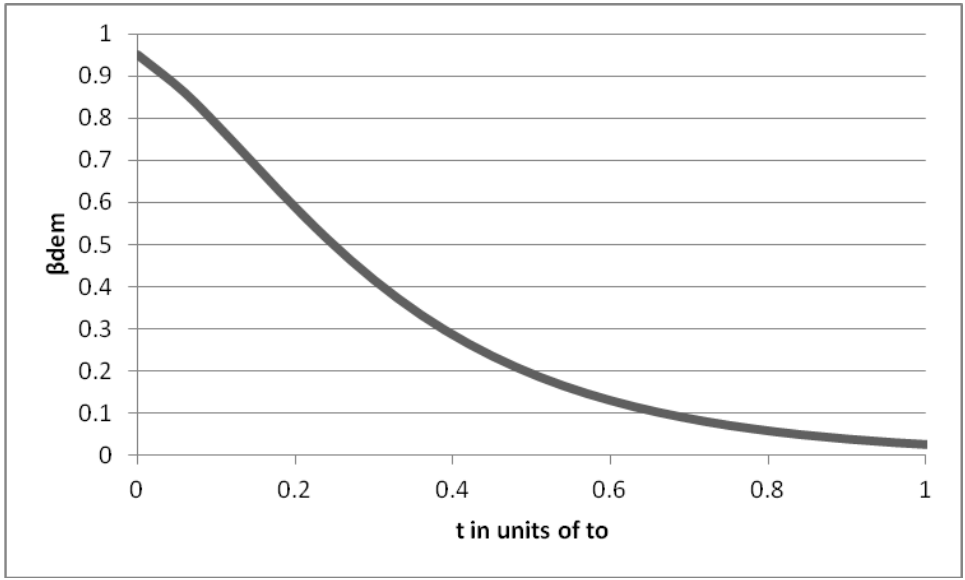

1c

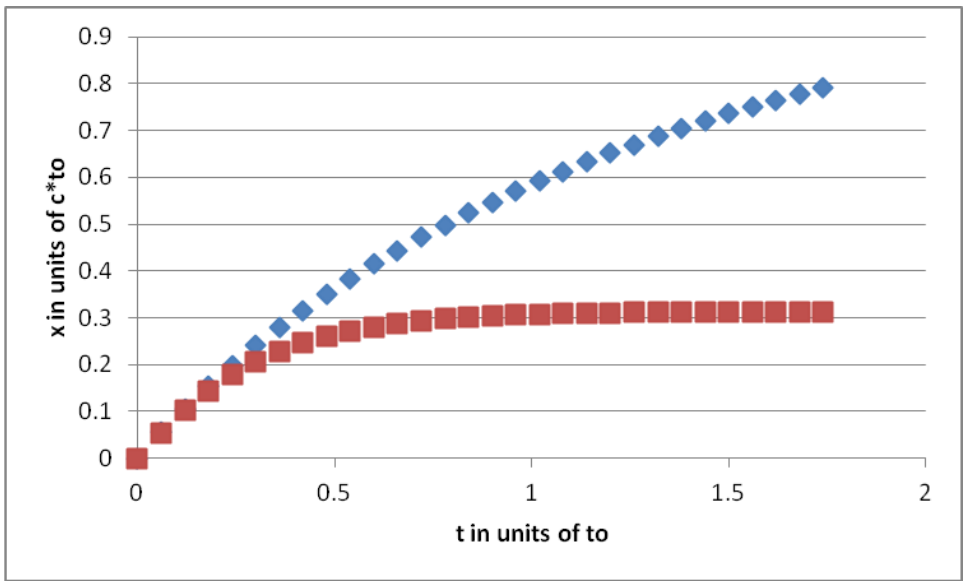

1d

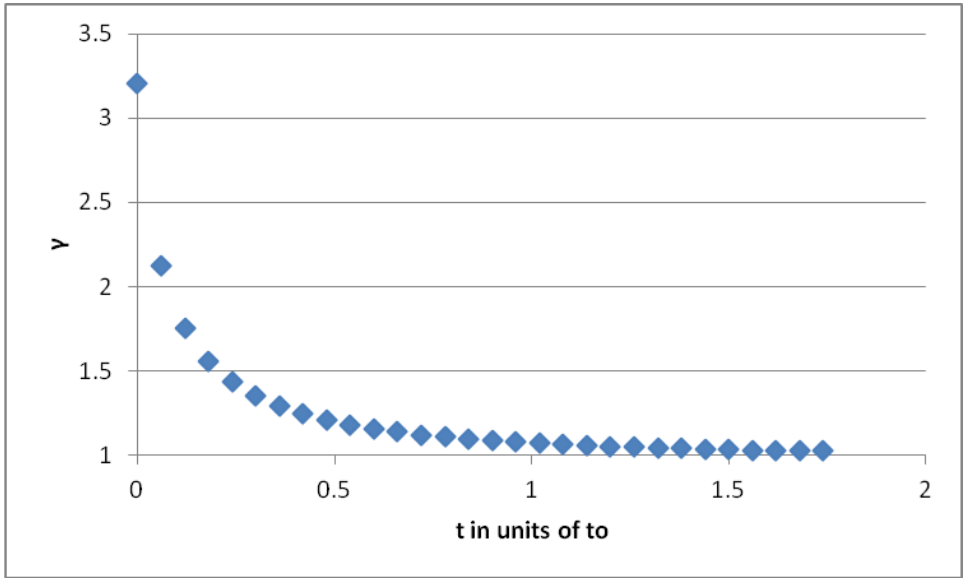

1e

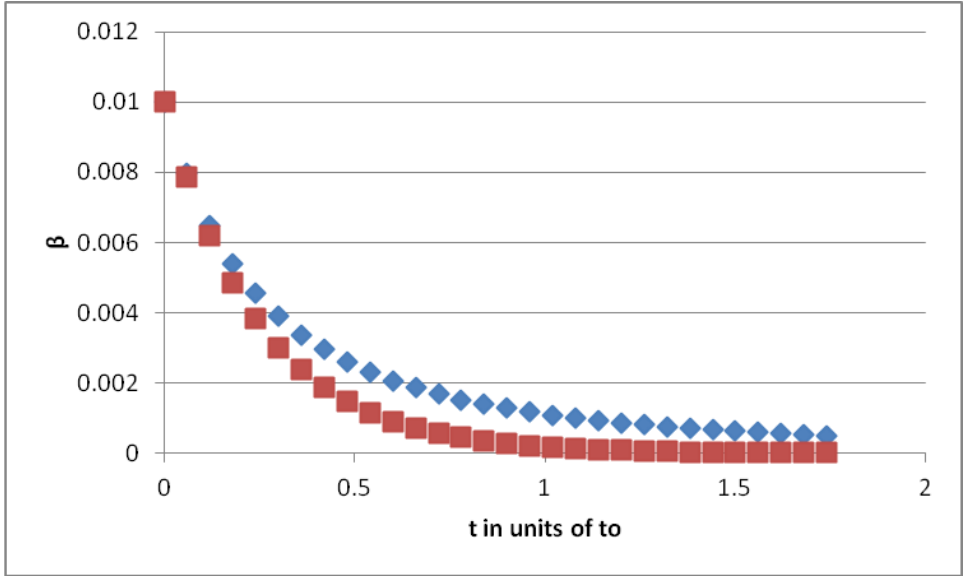

2a)

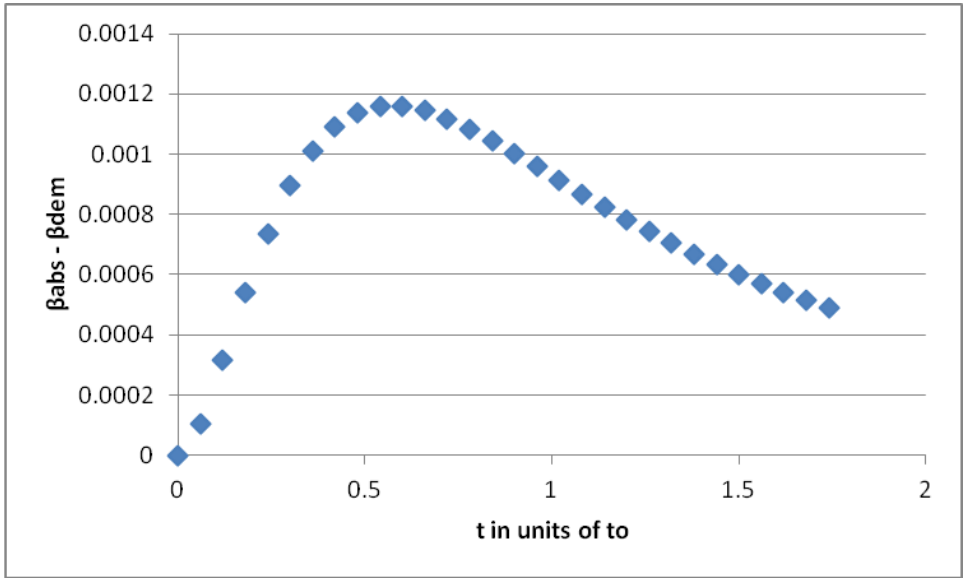

2b)

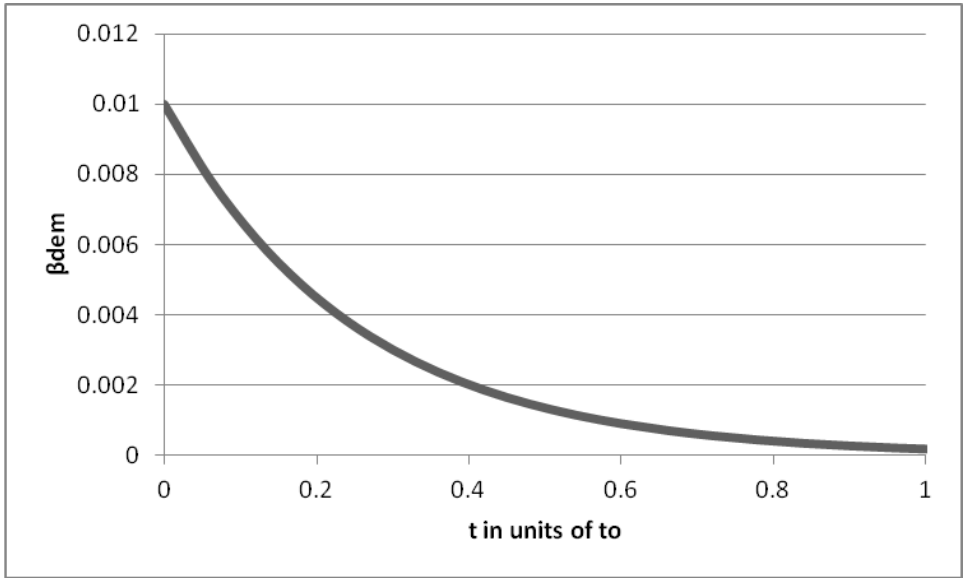

2c)

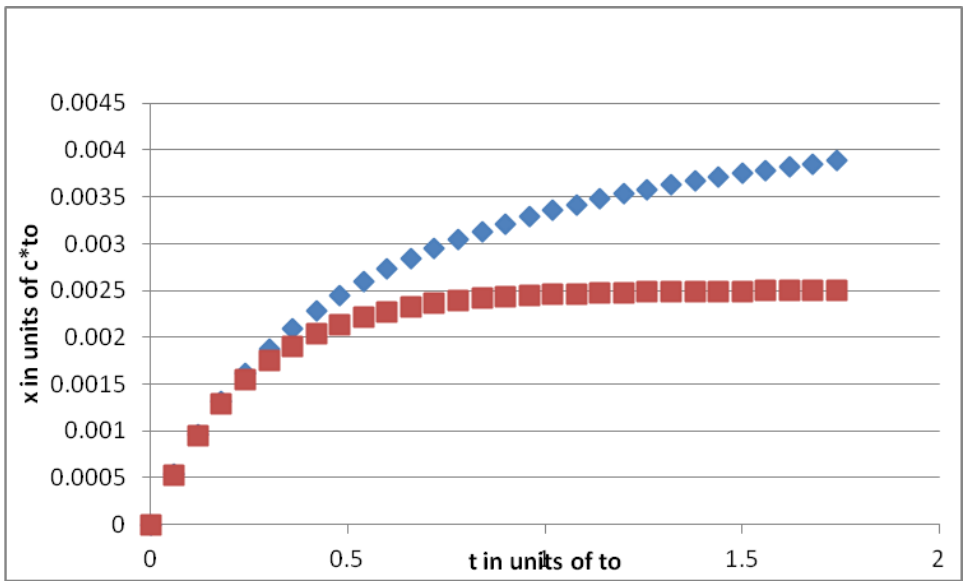

2d)